\documentclass[twocolumn]{aastex6}

\usepackage{graphicx}
\usepackage{rotating}
\usepackage{amssymb}
\usepackage{savesym}
\savesymbol{splitbox}
\usepackage{adjustbox}
\restoresymbol{adjustbox}{splitbox}
\begin{document}

\title{New Photometrically Variable Magnetic Chemically Peculiar Stars in the ASAS-3 Archive}

%% Use \author, \affil, plus the \and command to format author and affiliation 
%% information.  If done correctly the peer review system will be able to
%% automatically put the author and affiliation information from the manuscript
%% and save the corresponding author the trouble of entering it by hand.
%%
%% The \affil should be used to document primary affiliations and the
%% \altaffil should be used for secondary affiliations, titles, or email.

%% Authors with the same affiliation can be grouped in a single
\author{Stefan H{\"u}mmerich\altaffilmark{1}}
\affil{American Association of Variable Star Observers (AAVSO), Cambridge, USA}
\affil{Bundesdeutsche Arbeitsgemeinschaft f{\"u}r Ver{\"a}nderliche Sterne e.V. (BAV), Berlin, Germany}

\author{Ernst Paunzen}
\affil{Department of Theoretical Physics and Astrophysics, Masaryk University, Kotl\'a\v{r}sk\'a 2, 611 37 Brno, Czech Republic}

\and

\author{Klaus Bernhard}
\affil{American Association of Variable Star Observers (AAVSO), Cambridge, USA}
\affil{Bundesdeutsche Arbeitsgemeinschaft f{\"u}r Ver{\"a}nderliche Sterne e.V. (BAV), Berlin, Germany}


\altaffiltext{1}{ernham@rz-online.de}

\begin{abstract}

The magnetic Ap or CP2 stars are natural atomic and magnetic laboratories and ideal testing grounds for the evaluation of model atmospheres. CP2 stars exhibiting photometric variability are traditionally referred to as $\alpha^2$ Canum Venaticorum (ACV) variables. Strictly periodic changes are observed in the spectra and brightness of these stars, which allow the derivation of rotational periods. Related to this group of objects are the He-weak (CP4) and He-rich stars, some of which are also known to undergo brightness changes due to rotational modulation. Increasing the sample size of known rotational periods among CP2/4 stars is an important task, which will contribute to our understanding of these objects and their evolution in time. We have compiled an extensive target list of magnetic chemically peculiar (CP2/4) stars from the General Catalogue of Ap, HgMn, and Am stars. In addition to that, a systematic investigation of early-type (spectral types B/A) variable stars of undetermined type in the International Variable Star Index of the AAVSO (VSX) yielded additional ACV candidates, which were included in our sample. We investigated our sample stars using publicly available observations from the ASAS-3 archive. Our previous efforts in this respect led to the discovery of 323 variable stars in these data. Using a refined analysis approach, we were able to identify another 360 stars exhibiting photometric variability in the accuracy limit of the ASAS-3 data, thereby concluding our search for photometrically variable magnetic chemically peculiar stars in the ASAS-3 archive. Summary data, folded light curves and, if available, information from the literature are presented for all variable stars of our sample, which is composed of 334 bona-fide ACV variables, 23 ACV candidates and three eclipsing binary systems. Interesting and unusual objects are discussed in detail. In particular, we call attention to HD 66051 (V414 Pup), which was identified as an eclipsing binary system showing obvious rotational modulation of the light curve due to the presence of an ACV variable in the system.

\end{abstract}




\section{Introduction} \label{introduction}

Chemically peculiar (CP) stars, which comprise about 15\% of the upper main sequence stars between spectral types early B to early F, are characterized by abnormal line strengths of one or several elements. This peculiar abundance pattern is thought to be produced by selective processes (radiative levitation, gravitational settling) operating in calm radiative atmospheres \citep{2000ApJ...529..338R}. Evidence has accumulated that there is no clear boundary between normal and CP stars but rather a smooth transition in regard to peculiarity \citep{1987JApA....8..351L}. Furthermore, the CP phenomenon is not restricted to a particular evolutionary stage \citep{2006A&A...450..763K}. \citet{1974ARA&A..12..257P} divided the CP stars into the following four subgroups: CP1 stars (the metallic line or Am/Fm stars),  CP2  stars (the magnetic Bp/Ap stars), CP3 stars (the HgMn stars) and CP4 stars (the He-weak stars). Further groups of CP stars were subsequently defined, like e.g. the He-strong stars (\citealt{Berg1956,1970PASP...82..730M}) or the $\lambda$ Bootis stars (\citealt{1958SvA.....2..151P,2004IAUS..224..443P}).

The CP2 stars differ from the CP1 and CP3 objects in that they possess globally organized magnetic fields from about 300 G to several tens of kiloGauss (\citealt{2007A&A...475.1053A,2011IAUS..273..249K}), which also holds true for the CP4 objects. CP2 stars show a nonuniform distribution of chemical elements, which manifests itself in the formation of spots and patches of enhanced element abundance \citep{1981A&A...103..244M}, in which flux is redistributed through bound-free and bound-bound transitions \citep[e.g.][]{2013A&A...556A..18K}. Therefore, as the star rotates, strictly periodic changes are observed in the spectra and brightness of many CP2 stars, which are satisfactorily explained by the oblique rotator model \citep{1950MNRAS.110..395S}. CP2 stars exhibiting photometric variability are traditionally referred to as $\alpha^2$ Canum Venaticorum (ACV) variables \citep{Samu07}.

CP3 stars do not show strong large-scale organized magnetic fields, and the discussion about the presence of tangled magnetic fields is ongoing \citep[e.g.][]{2013A&A...554A..61K}. However, the line-profile variations detected in the spectra of these stars have also been interpreted in the terms of abundance inhomogeneities (\citealt{2002ApJ...575..449A,2006MNRAS.371.1953H}). Therefore, rotationally induced photometric variability at some level would be expected. While photometric variations in CP3 stars have been established beyond doubt, the underlying mechanism is still a matter of debate \citep[e.g.][]{2014A&A...561A..35M}. However, rotational modulation due to surface spots in CP3 stars is believed to produce only marginal photometric amplitudes \citep{2013MNRAS.429..119P}, which can likely be studied with high-precision (space) photometry only. In the preparatory stage of our investigation, some CP3 stars were checked for light variability in ASAS-3 data, albeit with null results, which substantiates this assumption. However, this question is beyond the scope of the present investigation, which concentrates on the classical magnetic CP2/4 stars.

CP2 stars are natural atomic and magnetic laboratories and, because of their unusual abundance patterns, ideal testing grounds for the evaluation of model atmospheres \citep{2009A&A...499..567K}. Increasing the sample of known CP2 stars is therefore an important task considerable effort has been devoted to in the past (\citealt{1986A&AS...64....9M,1998A&AS..133....1P,2011AA...525A..16P,2012MNRAS.420..757W}).

Our own efforts in this respect (\citealt{2015A&A...581A.138B,2015AN....336..981B}) have produced extensive lists of new ACV variables and candidates that have been found by an investigation of publicly available sky survey data. \citet[Paper 1 hereafter]{2015A&A...581A.138B} investigated the photometric variability of Ap stars using observations from the third phase of the All Sky Automated Survey \citep[ASAS-3,][]{2002AcA....52..397P} and identified 323 variable stars (mostly ACV variables), 246 of which were reported as variable objects for the first time. As an expansion of this work, we here report on the discovery of an additional 360 variable stars in ASAS-3 data which have been found by a refined analysis approach. In agreement with our expectations, the new sample is also composed mostly of bona-fide ACV variables.

Observations and target selection are described in Section \ref{observ}, data analysis and classification in Section \ref{dataan}. Results are presented and discussed in Section \ref{result}, and we conclude in Section \ref{conclu}.

\section{Observations and Target Stars} \label{observ}

\subsection{Characteristics of ASAS-3 data} \label{charac}

The aim of the All Sky Automated Survey (ASAS) is the detection and investigation of any kind of photometric variability. To this end, ASAS constantly monitored the entire southern sky and part of the northern sky to about $\delta$\,$<$\,+28\degr. The third phase of the project, ASAS-3, lasted from 2000 until 2009 \citep{2002AcA....52..397P}. The employed instrumentation, which was situated at the 10-inch astrograph dome of the Las Campanas Observatory in Chile, consisted of two wide-field telescopes equipped with f/2.8 200\,mm Minolta lenses and 2048 x 2048 AP 10 Apogee detectors that covered a field of sky of 8\fdg8x8\fdg8. About 10$^{7}$ sources brighter than $V \approx$ 14 mag were monitored in Johnson $V$. The achieved CCD resolution was about 14\farcs8 / pixel, which led to an astrometric accuracy of around 3 -- 5\arcsec\ for bright stars and up to 15.5\arcsec\ for fainter stars. As a result, photometry in crowded fields, as in star clusters, is rather uncertain. A field was typically observed every one, two or three days \citep{2014IAUS..301...31P}. This observing cadence results in strong daily aliasing and renders the interpretation of the resulting Fourier amplitude spectra ambiguous.

The ASAS-3 archive contains reasonable photometry for stars in the magnitude range 7\,$\la$\,$V$\,$\la$\,14. However, the most accurate data were obtained for targets in the magnitude range 8\,$\la$\,$V$\,$\la$\,10. Here, the typical scatter is about 0.01\,mag \citep[e.g.][]{2014IAUS..301...31P}. However, because of the long time base of almost ten years, ASAS-3 data allow for the detection of periodic signals with very small amplitudes. For instance, \citet{2014JAD....20....5D} identified periodic variables with a range of variability of 0.01 -- 0.02\,mag in the magnitude range of 7\,$\la$\,$V$\,$\la$\,10. \citet{2014IAUS..301...31P} estimated that periodic signals with amplitudes as low as about 5\,millimag (mmag) can be detected.

\begin{figure}
\begin{center}
\includegraphics[width=0.47\textwidth,natwidth=750,natheight=610]{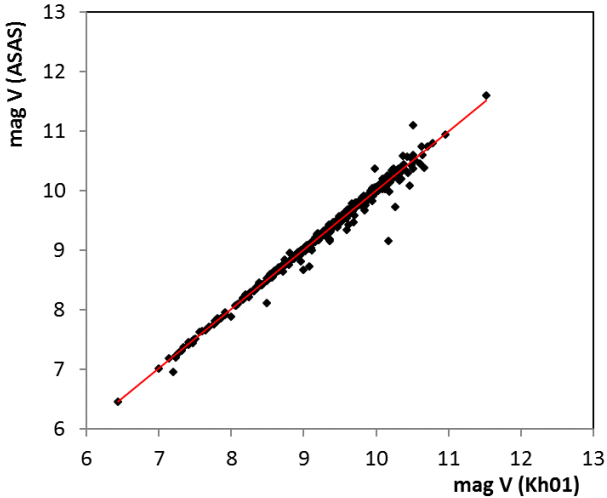}
\caption{Comparison between mean ASAS-3 $V$ magnitudes and $V$ magnitudes given by \citet{2001KFNT...17..409K} for the stars of our final sample.} 
\label{Fig_ASASTycho}
\end{center}
\end{figure}

\citet{2002AcA....52..397P} has shown that the zero-points of the ASAS-3 and Hipparcos photometry agree to within about $\pm$0.015 mag for stars lying close to the frame center. However, flat-fielding issues, missing color information and blending may result in much larger differences. We have investigated the agreement between mean ASAS-3 $V$ magnitudes and $V$ magnitudes given by \citet{2001KFNT...17..409K} for the stars of our final sample. The results are shown in Figure \ref{Fig_ASASTycho} and indicate very good general agreement between both sources. In most cases, unresolved close companions are responsible for the observed discrepancies.

\subsection{Target Stars} \label{target}

An intial list of target stars was created by selecting CP2 stars or CP2 star candidates and He-weak (CP4) / He-strong objects from the most recent version of the Catalogue of Ap, HgMn, and Am stars \citep[][RM09 hereafter]{2009A&A...498..961R}. Objects in the RM09 catalogue are not explicitly subdivided according to the classification established by \citet{1974ARA&A..12..257P}. We therefore resorted to the listed spectral types to distinguish between the different groups of CP stars (mainly denoted as ‘Si’, ‘Sr’, ‘Sr Eu Si’, ‘He weak’, ‘Hg Mn’, and so on). The resulting list of stars was cross-matched with the Tycho-2 catalogue \citep{2000A&A...355L..27H}; unlike our initial approach, where we defined a cut-off at $V_{\mathrm T}$\,$<$\,11\,mag (cf. Paper 1, section 2), no brightness limit was imposed.

In addition to that, a systematic investigation of early-type (spectral types B/A) variable stars of undetermined type in the AAVSO International Variable Star Index \citep[VSX;][]{2006SASS...25...47W} yielded additional ACV candidates, which were added to our target list.

We consulted the GCVS \citep{Samu07}, VSX, SIMBAD \citep{2000A&AS..143....9W} and VizieR \citep{2000A&AS..143...23O} databases in order to check for an entry in variability catalogues and to collect literature information on our target stars. Known ACV variables with well determined parameters were dropped from our sample; suspected or misclassified variables and variables of undetermined type were kept.

\section{Data Analysis and Classification} \label{dataan}

\subsection{Data Processing and Period Analysis}  \label{datapr}

The light curves of our sample stars were downloaded from the ASAS-3 website\footnote{http://www.astrouw.edu.pl/asas/}. For the present investigation, a refined analysis approach was developed with the intention of discovering variable objects that might have been missed by the imposed criteria in our previous work (Paper 1). In order to retain as many objects as possible, no lower limit was imposed on the number of observations in the ASAS-3 archive. In Paper 1, we restricted our analysis to promising candidates in order to keep our sample down to a manageable size. As promising candidates, we defined stars showing a larger scatter than usually observed for apparently constant stars in the corresponding magnitude range with comparable instruments \citep{2004AJ....128.1761H}. No such criteria were imposed in the present investigation; instead, every individual ASAS dataset was roughly cleaned of outliers and searched for periodic signals in the frequency domain of 0\,$<$\,$f$ (cycles per day; c/d hereafter)\,$<$\,20 using \textsc{Period04} \citep{2005CoAst.146...53L}.

Furthermore, the periodicity detection threshold was lowered significantly. Only objects with semi-amplitudes of $\gtrsim$ 0.007 mag (as derived with \textsc{Period04}) were considered in Paper 1. In the present investigation, all objects exhibiting variability with a semi-amplitude of at least 0.004 mag were subjected to a more detailed analysis. This limit is an experiental value based on our own extensive experience in dealing with the ASAS-3 data and the results of \citet{2014JAD....20....5D} and \citet{2014IAUS..301...31P}. It was chosen as a compromise between retaining variables with small amplitudes and eliminating spurious detections.

In the next step of analysis, left-over data points with a quality flag of 'D' (='worst data, probably useless') were rejected and remaining outliers were carefully removed by visual inspection. Furthermore, the data were checked for the presence of systematic trends. These were mostly due to strong blending effects which might result in significant additional scatter due to the inclusion of part of a neighbouring star's flux \citep{2014AcA....64..115S} or instrumental long-term trends that could introduce spurious signals into the data. Depending on the severity of artifacts, the affected datasets were either rejected or the trends were removed.

To refine the initial frequency analysis, the pretreated datasets were again searched for periodic signals in the frequency domain of 0\,$<$\,$f$\,(c/d)\,$<$\,10 with \textsc{Period04}. The data were folded on the resulting best fitting frequency and visually inspected. Objects exhibiting convincing phase plots were kept.

The light curves of CP2 stars can be well described by a sine wave and its first harmonic (\citealt{1984A&AS...55..259N,1985A&AS...60...17M,1987A&AS...70...33H,2015AN....336..981B}). We performed a least-squares fit to the data using \textsc{Period04}. Each light curve was fitted using a Fourier series consisting of the fundamental sine wave and its first harmonic, from which the light curve parameters (semi-amplitudes $A_{\mathrm 1}$, $A_{\mathrm 2}$, and the corresponding phases $\phi_{\mathrm 1}$, $\phi_{\mathrm 2}$) were derived.

As pointed out, the light curves of most ACV variables are sinusoidal. In orientations where two photometric spots of overabundant optically active elements come into view during a single rotation cycle, the light curve becomes a double wave \citep{1980A&A....89..230M}. If the two spots are of similar extent and photometric properties, the resulting 'maxima' will be of approximately the same height. Therefore, a twice longer (or shorter) rotation period cannot be excluded. This holds especially true for objects with very small amplitudes and/or significant scatter in their light curves.

In addition to that, alias periods cannot be totally excluded because of the strong daily aliasing inherent to ASAS-3 data (cf. section \ref{charac}). However, we have checked the period solution of all doubtful cases and are confident that we have come up with the period that fits ASAS-3 data best. This assumption is further corroborated by the generally very good agreement of our period solutions to those from the literature (cf. also Paper 1).

\subsection{Classification}  \label{classi}

For the final classification, all available information (spectral type, colour indices, period, shape of the light curve, Fourier amplitude spectrum) was taken into account. Except for the eclipsing binary systems (see below), all stars in our final sample exhibit a variability pattern that is in general accordance with rotational modulation caused by spots. HD 66051 (V414 Pup) is a special case in that it clearly shows both orbital (eclipses) and rotational modulation.

However, caution has to be taken, as it is not straightforward to distinguish between variability induced by rotation and other sources like e.g. pulsation or orbital motion. This holds especially true when analysing variable stars whose photometric amplitudes are near the detection limit of the employed data, as is the case for many of our targets.

\begin{figure}
\begin{center}
\includegraphics[width=0.47\textwidth,natwidth=555,natheight=394]{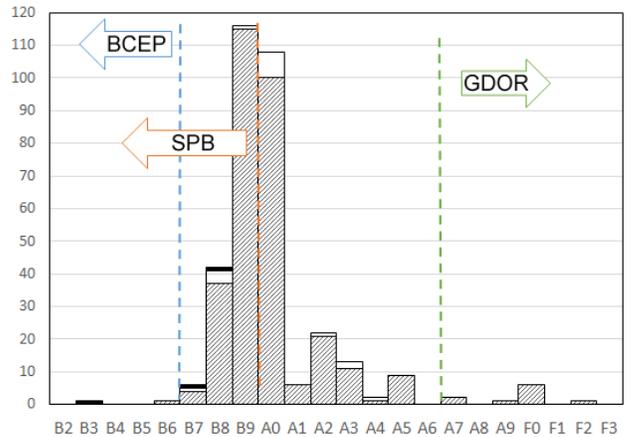}
\caption{Distribution of spectral types among the stars of our final sample. Only stars with accurate spectral classifications have been considered. Confirmed CP stars are indicated by the hatched area. The white and black areas denote, respectively, CP star candidates and CP4 / He-strong stars. The approximate loci of some important groups of early-type pulsating variables are indicated.} 
\label{distri}
\end{center}
\end{figure}

Pulsation as the underlying mechanism of the observed variability can be ruled out for most objects of our sample on the following grounds. The vast majority of our sample stars is found between spectral types B7 to A5 (see Figure \ref{distri}). Therefore, pulsators that are exclusively found among earlier spectral types, like e.g. $\beta$ Cephei variables (GCVS-type BCEP, spectral types $\sim$O8--B6), or primarily among later spectral types, as e.g. the $\gamma$ Doradus stars (GCVS-type GDOR, spectral types $\sim$A7--F7), are not expected to contribute much to our sample. Of course, inaccuracies / difficulties in spectral classification have to be considered, and the spectral types shown in Figure \ref{distri} might be uncertain by several subclasses.

Furthermore, our target stars exhibit photometric periods longer than $P >$ 0.5 days. We can thus exclude the presence of $\delta$ Scuti variables (GCVS-type DSCT) or other short period pulsators.

On the other hand, some types of pulsating variables partly overlap with ACV stars in respect to spectral type and period. The so-called slowly pulsating B (GCVS-type SPB) stars, for instance, are encountered down to spectral type B9 (Figure \ref{distri}) and exhibit periods between about 0.4 to 5 days. The $\gamma$ Doradus stars are found between spectral types A7 to F7; observed periods usually range from 0.3 to 3 days.

One way of distinguishing these types of variable stars is an investigation of their Fourier amplitude spectra. Many kinds of pulsators, like e.g. SPB and $\gamma$ Doradus stars, show multiple periods and quite different frequency spectra from rotating variables. For instance, harmonics of pulsation modes are only expected to be present in frequency spectra when the amplitude is large. On the other hand, harmonics are a consequence of localized spots and a characteristic of the frequency spectra of rotating variables \citep{2015MNRAS.451.1445B}.

Spots form and decay in late-type, active stars, and differential rotation might led to the presence of multiple, closely-spaced periods in these objects \citep{2015A&A...583A..65R}. However, the presence of starspots in stars with radiative envelopes, like B/A stars, is still a matter of some controversy. Recently, Balona and coworkers have collected evidence that A-type stars are active and show starspots in the same way as their cooler counterparts \citep[e.g.][]{2013MNRAS.431.2240B}. Nevertheless, the spots on CP2 stars are of a different nature (abundance patches) and constitute durable configurations that remain stable for decades, probably as a consequence of strong magnetic fields.

Differential rotation, however, plays an important role in A-type stars (\citealt{2012A&A...542A.116A,2013A&A...550A..94S,2016arXiv160407003B}). However, to our knowledge, no study of the possible effects of differential rotation on CP2 star light curves exists. Judging from the stability of the periods and light curves among ACV variables, we assume that these effects, if present, are generally small. However, it must not be dismissed that the presence of multiple, closely-spaced frequencies in the Fourier amplitude spectra of early-type stars might be due to differential rotation and need not automatically imply pulsation. Apart from this special scenario, however, the presence of multiple periods is not to be expected in CP2 stars and interpreted by us as an indication of pulsation in a non-CP2 star as the underlying mechanism of the observed photometric variability.

\begin{figure}
\begin{center}
\includegraphics[width=0.47\textwidth,natwidth=798,natheight=585]{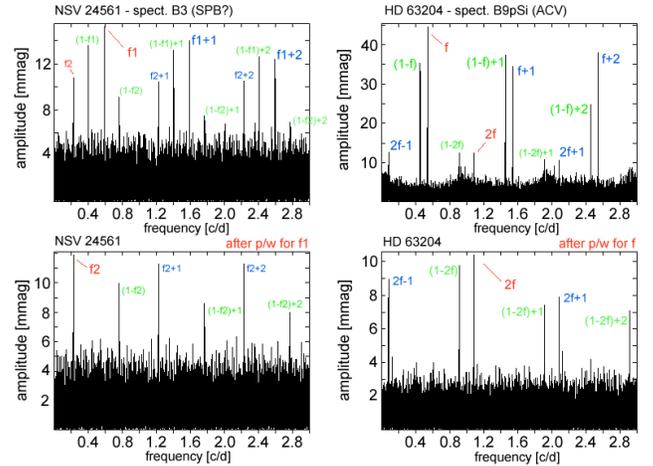}
\caption{Fourier amplitude spectra of a multi-periodic, pulsating variable (NSV 24561, spectral type B3, likely an SPB star; left panels) and a rotating variable (HD 63204, spectral type B9pSi, a confirmed ACV variable; right panels). The plots have been based on unwhitened ASAS-3 data (upper panels) and ASAS-3 data that have been prewhitened with, respectively, $f1$ and $f$ (lower panels). Significant frequencies are indicated in red. The corresponding $(f+1)$ and $(1-f)$ aliases are indicated, respectively, in blue and green.} 
\label{ACVSPB}
\end{center}
\end{figure}

Figure \ref{ACVSPB} shows the Fourier amplitude spectra of two B-type stars and illustrates the described differences in the frequency spectra between a multi-periodic, pulsating variable (NSV 24561, spectral type B3, likely an SPB star) and a rotating variable (HD 63204, spectral type B9pSi, a confirmed ACV variable, cf. \citealt{2015A&A...581A.138B}). No harmonics are seen in the spectrum of NSV 24561, which exhibits two significant low frequencies. In contrast, only one frequency and its first harmonic are present in the frequency spectrum of HD 63204.

It has to be kept in mind, though, that these assumptions are simplifications that do not represent Nature with all its intricacies. For instance, recent evidence from Kepler data indicates that rotational frequencies might possibly be present in SPB variables and result in the presence of harmonics in the corresponding Fourier amplitude spectra \citep{2015MNRAS.451.1445B}.

Finally, and most importantly, the spectra of all these kinds of pulsating variables are not characterized by the abnormal abundance patterns of the CP2/4 stars, which are a confirmed characteristic of most of our target stars and are expected for the CP2 star candidates in our sample. Generally, pulsation is not to be expected in CP2 stars. The only proven form of pulsational variability among this type of CP stars is observed in the so-called rapidly oscillating Ap (roAp) stars \citep{1982MNRAS.200..807K} which exhibit photometric variability in the period range of ~5-20 min (high-overtone, low-degree, and non-radial pulsation modes). This is very different from what has been observed for our sample stars. We therefore feel confident in ruling out pulsation as the underlying cause of the observed photometric variability in most of our sample stars.

The discrimination between rotational modulation and variability induced by orbital motion (as observed in ellipsoidal variables, GCVS-type ELL, and eclipsing binaries) is more difficult, though. Generally, it is not possible to distinguish between both types of variations without additional spectroscopic information \citep[Paper 1;][]{2015MNRAS.451.1445B}. For some CP stars, for instance, a double-wave structure of the photometric light curve has been observed \citep{1980A&A....89..230M}. The light curves of these 'double-humped' ACVs are not to be distiguished from the light curves of ellipsoidal variables on grounds of single-passband photometric data alone. However, it has been shown that the incidence of ellipsoidal or eclipsing variables among CP2 stars is very low (\citealt{1985A&A...146..341G,2004ASPC..318..297N,2014MNRAS.440L...6H}) and even in a sample of some hundred stars, only very few ellipsoidal or eclipsing variable star candidates are to be expected (cf. Paper 1).

We have identified three eclipsing binary stars among our targets (CPD-20 1640, HD 66051, and HD 149334, cf. section \ref{result}). While the system of HD 66051 (V414 Pup) definitely hosts a CP2 star, the CP classification of the other two objects is doubtful; thus, spectroscopic investigations are needed to confirm or reject the assumed presence of a CP2 star in these systems. Furthermore, with the available data, we are not able to distinguish between rotational and orbital modulation as the underlying cause of the photometric variability of the CP4 star HD 161733A. Section \ref{noteso} provides a detailed discussion of these objects.

Before the background of the characteristic light variations and Fourier amplitude spectra, the confirmed CP2/4 nature of our targets, and the available spectral classifications, which go along well with the observed colour indices, the most likely explanation of the observed light variations in the majority of our sample stars is the redistribution of flux in spots of overabundant optically active chemical elements. We are therefore confident that most of the confirmed CP2/4 stars in our sample are bona-fide ACV variables ($N = 334$). The few exceptions or special cases are commented on in section \ref{noteso}.

This assumption is further corroborated by the fact that (with the exception of the eclipsing binaries) all light curves of our sample stars can be well represented by a sine wave and its first harmonic –- a procedure which has been shown to adequately describe the light curves of ACV variables (cf. section \ref{datapr}). The photometrically variable CP2 star candidates in our sample are here proposed as ACV variable candidates (type ACV: in Table \ref{table_master}; $N = 23$) on grounds of their periods and typical photometric variability but need spectroscopic confirmation of their CP status. The two CP4 objects and the He-strong star are also designated as ACV candidates, as other mechanisms beside rotational modulation might be at work in these objects (section \ref{noteso}).

\section{Results} \label{result}

\subsection{Presentation of Results}  \label{presen}

Employing the methodology outlined above (sections \ref{observ} and \ref{dataan}), 360 stars exhibiting photometric variability in the accuracy limit of the ASAS-3 data were identified among the stars of our target list. We have ruled out pulsation as the underlying mechanism of the observed variability in most of our targets and are confident of the applicability of our classifications. Table \ref{table_overvi} gives statistical information on the composition of the final sample.

\begin{table}
\caption{Statistical information on the composition of the final sample.}
\label{table_overvi}
\begin{center}
\begin{tabular}{lr}
\hline
\hline 
Type & Number of objects \\
\hline
ACV variables  & 334 \\
ACV variable candidates  &  23 \\
eclipsing binary systems & 3 \\
total of variable stars & 360 \\
\hline
\end{tabular}
\end{center}
\end{table}

We have searched the SIMBAD, VizieR, and VSX databases for previously published information on our targets. According to these sources, most of the investigated CP stars have never been the subject of a light variability analysis before and are here presented as variable stars for the first time. Some of our target stars have been previously investigated and found constant or probably constant; other objects have been identified as variable stars with or without a given period in the literature, but their variability types have not been determined or they have been misclassified.

Table \ref{table_master} presents essential data and light curve fit parameters for our sample stars and is organized as follows:
\begin{itemize}
\item Column 1: star name, HD number, or other conventional identification.
\item Column 2: identification number from RM09.
\item Column 3: right ascension (J2000; Tycho-2).
\item Column 4: declination (J2000; Tycho-2).
\item Column 5: variability type, according to GCVS convention (ACV / ACV: / EA). 
\item Column 6: $V$ magnitude range, as derived from the Fourier fit to the ASAS-3 data.
\item Column 7: period (d).
\item Column 8: epoch (HJD-2450000); time of maximum is indicated for ACV variables or candidates, time of minimum for the eclipsing binary systems.
\item Column 9: Semi-amplitude of the fundamental variation ($A_{\mathrm 1}$).\footnote{Columns 8--11 have only been calculated for ACV variables or candidates. In the case of HD 66051 (V414 Pup), the corresponding values have been calculated from a fit to the out-of-eclipse, rotationally induced variability (cf. section \ref{HD 66051}).}
\item Column 10: Semi-amplitude of the first harmonic variation ($A_{\mathrm 2}$).
\item Column 11: Phase of the fundamental variation ($\phi_{\mathrm 1}$).\footnote{The calculation of the phase values has been based on the times of observations as provided by the ASAS-3 database, i.e. HJD-2450000.}
\item Column 12: Phase of the first harmonic variation ($\phi_{\mathrm 2}$).
\item Column 13: Spectral classification, as listed in RM09; it is noteworthy that, as in the original catalogue, the 'p' denoting peculiarity has been omitted from the spectral classifications taken from RM09. For the five stars not included in this catalogue, the spectral types have been gleaned from the VSX and verified using the VizieR and SIMBAD catalogue services.
\item Column 14: $(B−-V)$ index, taken from \citet{2001KFNT...17..409K}.
\item Column 15: $(J−-K_{\mathrm s})$ index, as derived from the 2MASS catalogue \citep{2006AJ....131.1163S}.
\end{itemize}

The light curves of all objects, folded with the periods listed in Table \ref{table_master}, are presented in the Appendix (Figure 8).\footnote{Only part of Figure 8 is included; the complete figure is available from the authors.} Information from the literature and miscellaneous remarks on individual objects are listed in Table \ref{table_remarks}, which is organized as follows:

\begin{itemize}
\item Column 1: star name, HD number, or other conventional identification.
\item Column 2: variable star designation from the literature.
\item Column 3: variable star type from the literature.
\item Column 4: period (d) from the literature.
\item Column 5: period (d) from this work.
\item Column 6: reference in which, to the best of our knowledge, the object has been announced as a variable star for the first time.
\item Column 7: remarks / comments of a miscellaneous nature; an asterisk denotes stars, whose status as chemically peculiar objects is doubtful according to RM09.
\end{itemize}

Both tables are listed in their entirety in the Appendix. We are currently working on a statistical paper on the properties of ACV variables, which will include results from the literature as well as our own investigations (Paper 1, \citealt{2015AN....336..981B}, this paper). Therefore, in the present work, we restrict ourselves to the discussion of interesting and unusual objects.

\begin{table*}
\centering
\caption{Essential data for the stars identified as photometrically variable chemically peculiar stars or candidates in the present paper. Only part of the table is printed here for guidance regarding its form and content. The complete table is given in the Appendix.}
\label{table_master}
\begin{adjustbox}{max width=\textwidth}
\begin{tabular}{llcclcccccccccc}
\hline 
Star & ID (RM09) & $\alpha$ (J2000) & $\delta$ (J2000) & Type & Range ($V$) & Period & Epoch (HJD) & $A_{\mathrm 1}$ & $A_{\mathrm 2}$ & $\phi_{\mathrm 1}$ & $\phi_{\mathrm 2}$ & Spectral type & $(B-V)$ & $(J-K_{\mathrm s})$ \\
     &           &                  &                  &      & [mag]       & [d]    & [d]  & [mag]           & [mag]           & [rad]             & [rad]               &              & [mag]   & [mag] \\
\hline
HD 2957	          & 	760	&	00 32 44.08	&	-13 29 13.5	&	ACV	&	8.48-8.51 	&	4.6327(3)       &	4627.91(9)   &	0.011	&	0.002	&	0.846	&	0.750	&	B9 Cr Eu	&	+0.052	&	-0.017 \\
HD 3885	          &	1050	&	00 41 18.31	&	-19 51 45.9	&	ACV	&	9.79-9.81	&	1.81508(4) 	&	2910.72(4)   &	0.007	&	0.009	&	0.348	&	0.449	&	B9 Si	&	-0.090	&	-0.079 \\
HD 5823	          &	1540	&	00 59 39.29	&	-11 56 01.2	&	ACV	&	9.95-9.97	&	1.24520(2) 	&	3765.53(2)   &	0.007	&	0.004	&	0.723	&	0.646	&	F2 Sr Eu Cr	&	+0.304	&	+0.176 \\
HD 8783	          &	2110	&	01 24 00.43	&	-72 19 27.9	&	ACV	&	7.80-7.82	&	19.396(5)   	&	3794.5(4)    &	0.009	&	0.001	&	0.102	&	0.715	&	A2 Sr Eu Cr	&	+0.150	&	+0.039 \\
HD 12559	  &		&	02 03 29.36     &	+18 19 39.4	&	ACV:    &	8.41-8.44  	&	4.0358(3)       &	3338.62(8)   &	0.010	&	0.012	&	0.479	&	0.293	&	A2	&	+0.092	&	-0.034 \\
HD 16145	  &	4060	&	02 35 04.20	&	-17 17 22.3	&	ACV	&	7.64-7.67  	&	2.23766(7)      &	4525.52(4)   &	0.011	&	0.002	&	0.342	&	0.873	&	A0 Cr Sr Eu	&	+0.117	&	-0.023 \\
HD 20505	  &	5120	&	03 15 03.18	&	-59 11 36.1	&	ACV	&	9.88-9.90	&	2.04401(5) 	&	2666.49(4)   &	0.009	&	0.002	&	0.222	&	0.726	&	A2 Cr Sr	&	+0.120	&	+0.001 \\
HD 22032	  &	5540	&	03 33 11.64	&	+04 40 13.3	&	ACV	&	9.05-9.07	&	4.8589(3)   	&	3031.57(9)   &	0.010	&	0.001	&	0.839	&	0.523	&	A3 Sr Eu Cr	&	+0.455	&	+0.056 \\
HD 23509	  &	6020	&	03 41 29.82     &	-66 28 37.9	&	ACV:    &	7.75-7.78	&	1.48786(3) 	&	3010.64(3)   &	0.012	&	0.002	&	0.302	&	0.426	&	A3	&	+0.309	&	+0.167 \\
HD 27210	  &	6960	&	04 15 22.02	&	-51 44 27.1	&	ACV:    &	10.09-10.12	&	1.01438(1) 	&	3447.55(2)   &	0.015	&	0.000	&	0.120	&	--	&	A0	&	+0.064	&	-0.046 \\
HD 28238	  &	7210	&	04 27 35.96	&	+06 36 43.2	&	ACV	&	9.11-9.12	&	24.743(7)       &	2676.5(5)    &	0.007	&	0.001	&	0.615	&	0.177	&	A0 Sr Cr Eu	&	+0.241	&	+0.053 \\
HD 30374  	  &	7830	&	04 39 01.57	&	-75 06 10.1	&	ACV	&	10.02-10.04	&	1.55631(3) 	&	3018.65(3)   &	0.008	&	0.003	&	0.097	&	0.580	&	A0 Sr Eu Cr	&	+0.105	&	+0.081 \\
HD 240563	  &	8310	&	05 04 57.34	&	+08 50 05.6	&	ACV	&	10.09-10.12	&	2.9447(1)   	&	4399.78(6)   &	0.015	&	0.003	&	0.646	&	0.601	&	A3 Sr	&	+0.216	&	+0.078 \\
HD 245155	  &	9662	&	05 35 42.79	&	+25 16 29.4	&	ACV	&	9.67-9.69	&	0.705370(7)	&	3730.62(1) &	0.009	&	0.001	&	0.809	&	0.251	&	B9 Si Sr	&	+0.080	&	+0.109 \\
HD 38417	  &	10330	&	05 38 55.39	&	-71 38 20.1	&	ACV	&	9.61-9.63	&	2.16619(6)      &	3086.56(4)   &	0.002	&	0.008	&	0.096	&	0.011	&	A0 Sr	&	+0.121	&	+0.044 \\
HD 38912	  &	10450	&	05 49 13.10	&	+01 27 30.2	&	ACV	&	9.46-9.49  	&	1.46279(2) 	&	2539.74(3)   &	0.013	&	0.003	&	0.608	&	0.148	&	B8 Si	&	+0.266	&	+0.106 \\
HD 39082	  &	10500	&	05 50 23.85	&	+04 57 24.3	&	ACV	&	7.41-7.42  	&	0.764776(7)	&	4877.58(2) &	0.007	&	0.001	&	0.045	&	0.341	&	B9 Sr Cr Eu	&	+0.039	&	-0.041 \\
HD 40071	  &	10680	&	05 56 06.12	&	-13 08 07.2	&	ACV	&	8.06-8.08	&	1.98735(5) 	&	4477.82(4)   &	0.005	&	0.005	&	0.744	&	0.369	&	B9 Si	&	-0.042	&	-0.062 \\
HD 40383	  &	10780	&	05 58 37.50	&	+04 29 33.6	&	ACV	&	8.98-9.00	&	4.0364(3)       &	3644.86(8)   &	0.009	&	0.007	&	0.739	&	0.771	&	B9 Si	&	+0.215	&	+0.059 \\
HD 40678	  &	10876	&	06 01 29.21	&	+23 42 14.2	&	ACV	&	7.37-7.39	&	22.029(6)       &	2678.2(4)    &	0.012	&	0.001	&	0.238	&	0.541	&	A0 Si Sr	&	+0.158	&	-0.085 \\
                                                                                                                                                      
\end{tabular}                                                                                                                                                                 
\end{adjustbox}                                                                                                                                         
\end{table*}

\begin{table*}[ht]
\caption{Relevant information on single objects from the literature and miscellaneous remarks. An asterisk in column 7 ('Remarks/comments') denotes stars whose status as chemically peculiar objects is doubtful according to RM09. The following abbreviations are employed in column 7: R12 = \citet{2012MNRAS.427.2917R}; W12 = \citet{2012MNRAS.420..757W}. Only part of the table is printed here for guidance regarding its form and content. The complete table is given in the Appendix.}
\label{table_remarks}
\scriptsize{
\begin{center}
\begin{adjustbox}{max width=\textwidth}
\begin{tabular}{llccccl}
\hline
\hline
Star &  Var. desig. & Var. type &       Period (d)  &   Period (d)      &       Reference       &       Remarks/comments        \\
     &  Literature       & Literature     & Literature  & This work   &           & Literature \\      
\hline
HD 2957		&				&		&						&	4.6327(3)  	&		&	Null result for roAp pulsations \citep{1994MNRAS.271..129M}. \\
HD 3885		&	NSV 15149		&	VAR	&						&	1.81508(4) 	&	\citet{1982AAS...49..427W}	&	\\
HD 5823		&				&		&						&	1.24520(2) 	&		&	Null result for roAp pulsations \citep{2013MNRAS.431.2808K}. \\
HD 8783		&	SMC V2339		&	VAR:	&	21(or 1.05)? (GCVS)			&	19.396(5)  	&	GCVS	&	Non-member of the SMC according to the GCVS. The given period \\
		& 				& 		& 						& 	           	& 		& values are derived from variations of the peculiarity index ?a. \\
HD 12559	&	HIP 9602		&	VAR	&	2.01833 (VSX);  2.01837 (R12)		&	4.0358(3)  	&	\citet{2002MNRAS.331...45K}	&	R12: SPB/ACV (prob: 0.24/0.51) \\
HD 16145	&				&		&	2.24(or 4.47)? (RM09)			&	2.23766(7) 	&		&	Null result for roAp pulsations \citep{1994MNRAS.271..129M}. \\
HD 23509	&	HIP 17239		&	VAR	&	1.48779 (VSX); 1.48765 (R12)		&	1.48786(3) 	&	\citet{2002MNRAS.331...45K}	&	R12: SPB/DSCTC (prob: 0.23/0.33) \\
HD 27210	&				&		&						&	1.01438(1) 	&		&	* \\
HD 240563	&				&		&						&	2.9447(1)  	&		&	* \\
HD 245155	&				&		&						&	0.705370(7)	&		&	Constant or quality of data prevented detection (W12). \\
HD 38417	&				&		&						&	2.16619(6) 	&		&	* \\
HD 39082	&	NSV 16689		&	VAR 	&	0.76484 (VSX)				&	0.764776(7) 	&	\citet{1979AAS...36..477V}	&	 \\
HD 40678	&				&		&						&	22.029(6)  	&		&	Constant or quality of data prevented detection (W12). \\
BD-06 1402	&	HIP 28864 		&	VAR 	&	1.13065 (VSX)				&	1.13063(2) 	&	\citet{2002MNRAS.331...45K}	&	R12: EB/SPB (prob: 0.15/0.57)\\
		& ASAS J060540-0603.2		& DCEP-FU/MISC 	& 1.13058 (R12) 				& 	           	& & \citet{2012ApJS..203...32R}: ACV (prob: 0.8109) \\
HD 41869	&				&		&						&	5.2350(4)  	&		&	Constant or probably constant, blend? in STEREO data (W12). \\
HD 46105	&	NSV 16891		&	VAR	&						&	0.793263(8)	&	\citet{1982AAS...48..503R}	&	 \\
HD 46649	&				&		&						&	4.1792(3)  	&		&	RM09: A0 Si Cr ? * \\
HD 49797	&				&		&						&	1.22649(2) 	&		&	* \\
HD 50031	&				&		&						&	2.8743(1)  	&		&	Null result for roAp pulsations \citep{1994MNRAS.271..129M}. \\
HD 51342	&				&		&						&	1.43601(2) 	&		&	* \\
\hline
\end{tabular}
\end{adjustbox}
\end{center}
}
\end{table*}

\subsection{Notes on Individual Objects} \label{noteso}

The following sections contain notes and literature information on several unusual and interesting objects.

\subsubsection{CPD-20 1640 = NGC 2287 40}  \label{CPD-20 1640}
CPD-20 1640 is listed with a spectral type of A5pSiSr in the RM09 catalogue. It has been identified with Cox 40 (=NGC 2287 40) and is likely a member of the intermediate-age open cluster NGC 2287 \citep{2007A&A...470..685L}. However, its status as a CP2 star needs confirmation. While the listed spectral type is in accordance with this classification, it is not supported by the measured $\Delta a$ and $Z$ values and the non-detection of a magnetic field \citep[cf.][and references therein]{2007A&A...470..685L}.

\begin{figure}
\begin{center}
\includegraphics[width=0.47\textwidth,natwidth=979,natheight=801]{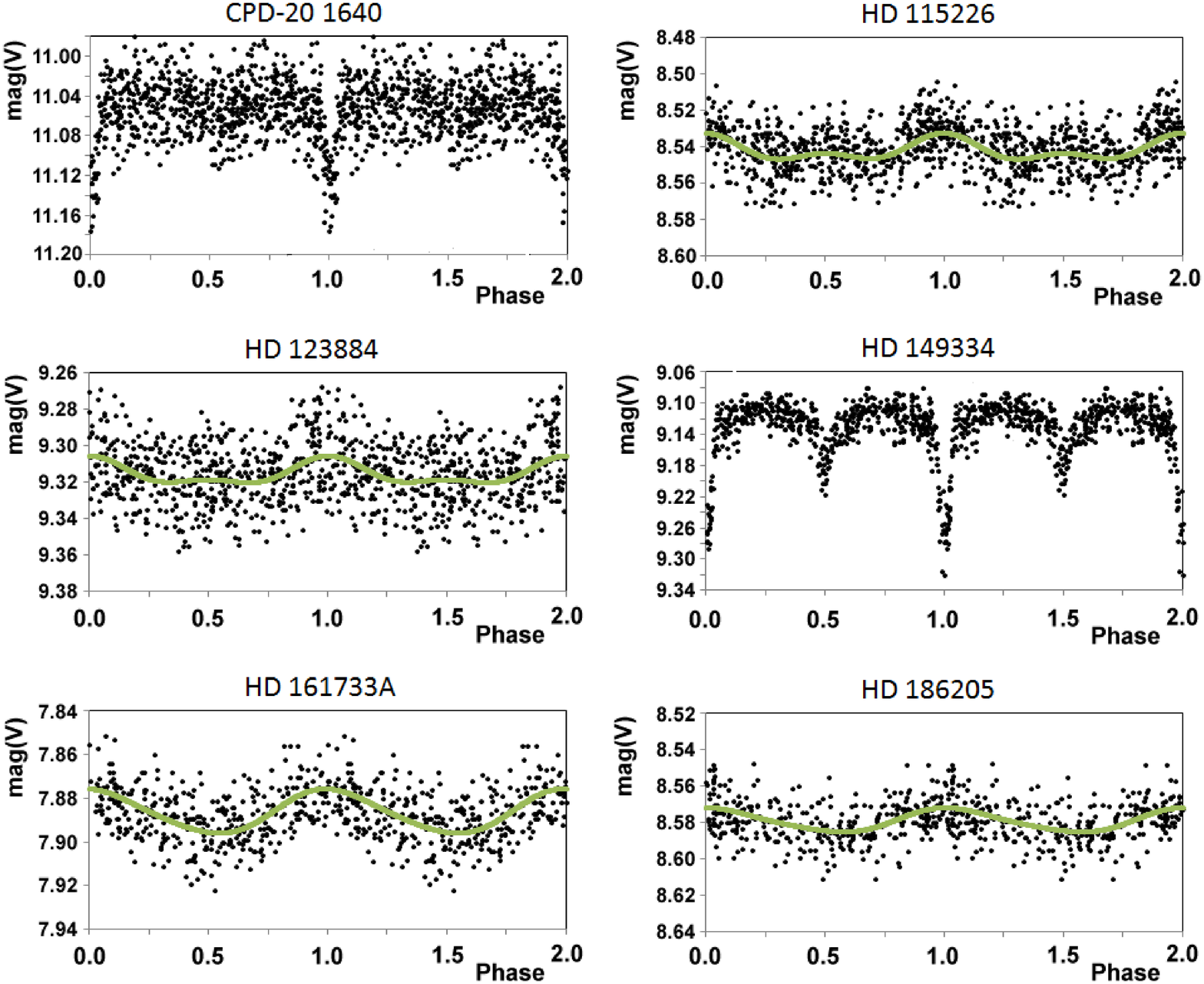}
\caption{Phase plots of objects discussed in Section \ref{noteso}, based on ASAS-3 data and folded with the periods given in Table \ref{table_master}.} 
\label{indobj}
\end{center}
\end{figure}

ASAS-3 data indicate that CPD-20 1640 is an eclipsing binary with a period of $P$ = 2.43400(2) d. The primary minimum is sharp and suggestive of a detached or semi-detached system (cf. Figure \ref{indobj}). If proven that at least one component of this system is indeed a classical CP2 star, CPD-20 1640 would be of high interest, as the incidence of CP2 stars in eclipsing binaries is very low (cf. section \ref{classi}). RM09 list only five candidates for eclipsing CP2 stars; only one (AO Velorum) has been confirmed \citep{2006A&A...449..327G}. Another confirmed eclipsing CP2 star (HD 66051) is presented in the following section. One good candidate (HD 70817) and three possible candidates were identified in Paper 1 but need spectroscopic confirmation. We therefore strongly encourage further studies of CPD-20 1640 in order to confirm or reject the assumed presence of a CP2 star in the system.

\subsubsection{HD 66051 = V414 Pup}  \label{HD 66051}
HD 66051 is a confirmed CP2 star of the Silicon subgroup (\citealt{1973AJ.....78..687B,1988mcts.book.....H}) and listed with a spectral type of A0pSi in the RM09 catalogue. Its photometric variability was discovered in Hipparcos data (HIP 39229; \citealt{1997AA...323L..61V}). The star was subsequently included in the GCVS as an ACV candidate (type ACV:); no period or epoch were given.

The star was discovered to be an eclipsing binary of Algol-type (GCVS-type EA) by \citet{2003IBVS.5480....1O}, who derived an orbital period of $P$ = 4.74922 d and a magnitude range of 8.79--9.12 mag ($V$) from a combination of Hipparcos and ASAS-3 data. Furthermore, the star was shown to exhibit additional variability with an amplitude of 0.05 mag and the same period which was interpreted as being due to 'ACV variations', i.e. rotationally induced variability caused by surface inhomogeneities on (at least) one of the system's components.

\begin{figure}
\begin{center}
\includegraphics[width=0.47\textwidth,natwidth=770,natheight=666]{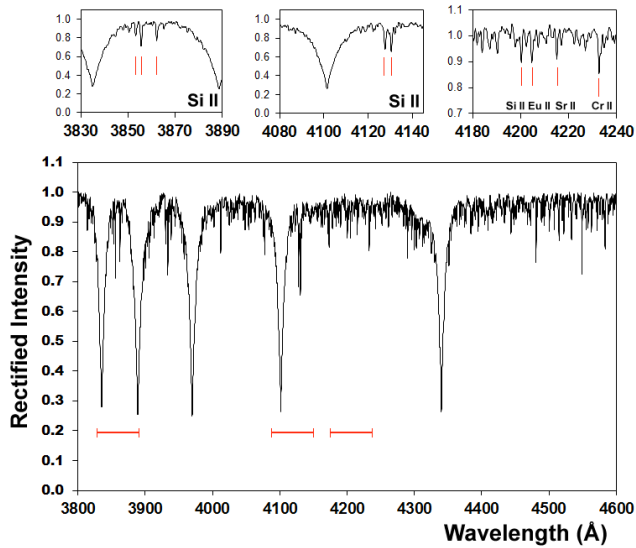}
\caption{Blue-violet spectral region of HD 66051 (V414 Pup), based on the HARPS spectrum obtained by \citet{2007A&A...470.1201D}. The upper panels show detailed views of regions of interest, which are indicated by the red lines in the lower panel. Some prominent lines are indicated.}
\label{Fig_HD66051_spec}
\end{center}
\end{figure}

No further detailed studies of HD 66051 exist in the VizieR and SIMBAD databases. However, a high resolution spectrum of the star is available in the archive of the 'Variable Star One-shot Project' \citep{2007A&A...470.1201D}, which was taken with the HARPS instrument \citep{2003Msngr.114...20M} at the ESO La Silla Observatory in Chile. Details on the spectroscopic observation can be found in \citet{2007A&A...470.1201D}. The spectrum confirms that HD 66051 harbours a CP2 Si star (cf. Fig. \ref{Fig_HD66051_spec}). In addition to that, enhanced lines of other elements, like e.g. Sr, Cr, and Eu, are present. The spectrum was obtained on JD 2453827.518802, which -- assuming the epoch of 2452167.867 as orbital phase $\varphi$ = 0 \citep{2003IBVS.5480....1O} -- corresponds to $\varphi_\mathrm{orb}$ = 0.46. The spectrum does not confirm the proposed SB2 nature of the system \citep{2007A&A...470.1201D}, which is likely due to coverage at an 'unfortunate' orbital phase.

\begin{figure}
\begin{center}
\includegraphics[width=0.47\textwidth,natwidth=485,natheight=337]{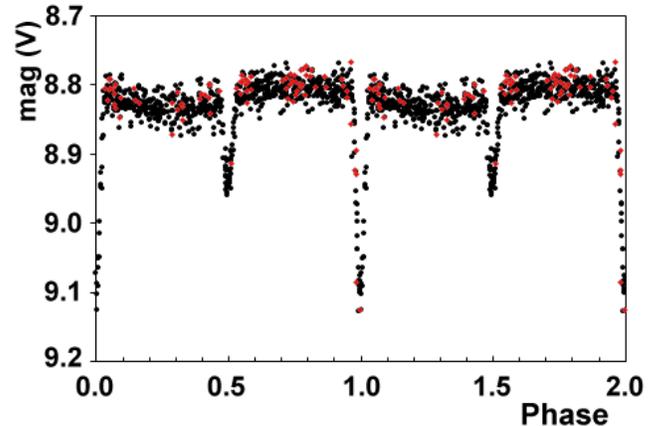}
\caption{Phase plot of HD 66051 (V414 Pup), based on Hipparcos data (red diamonds) and ASAS-3 data (black dots) and folded with $P$ = 4.74922 d. Hipparcos data have been transformed to the $V$ scale following \citet{2003IBVS.5480....1O}. Note the secondary variability due to the presence of a synchronously rotating ACV variable in the system. See text for details.}
\label{Fig_HD66051}
\end{center}
\end{figure}

We have investigated the object using all available data and confirm the findings of \citet{2003IBVS.5480....1O}. The longevity of the observed secondary variability in the light curve, which remains stable during the $\sim$9 years of ASAS-3 coverage (Fig. \ref{Fig_HD66051}), might be interpreted in terms of synchronous rotation, i.e. both stars are tidally locked.

HD 66051 is of great astrophysical interest. Firstly, the incidence of eclipsing binaries among CP2 stars is very low (cf. section \ref{classi}). Secondly, the system is quite unique in exhibiting both eclipses and obvious rotational variability due to abundance inhomogeneities, which opens up a lot of interesting possibilities for future research. We have already embarked on a detailed study of this star, the results of which will be presented in a future publication.

\subsubsection{HD 115226}  \label{HD 115226}
Using time-series spectroscopy, \citet{2008AA...479L..29K} identified HD 115226 as a rapidly oscillating Ap (roAp) star and inferred a pulsation period of 10.86 min from radial velocity variations in Pr \textsc{III}, Nd \textsc{III}, Dy \textsc{III} and the narrow cores of the hydrogen lines. They found $v_\mathrm{e}\,sin\,i$ = 25-30 km\textsuperscript{-1} and deduced a rotational period of $P_\mathrm{rot} \leq 3.0-3.5$ d. Furthermore, they established the presence of surface abundance inhomogeneities but did not detect any significant variability in the then available data from the ASAS-3 survey. However, marginal variability with $P$ = 3.61 d and an amplitude below 0.01 mag was detected in Hipparcos data \citep{1997AA...323L..61V}.

\begin{figure}
\begin{center}
\includegraphics[width=0.41\textwidth,natwidth=627,natheight=453]{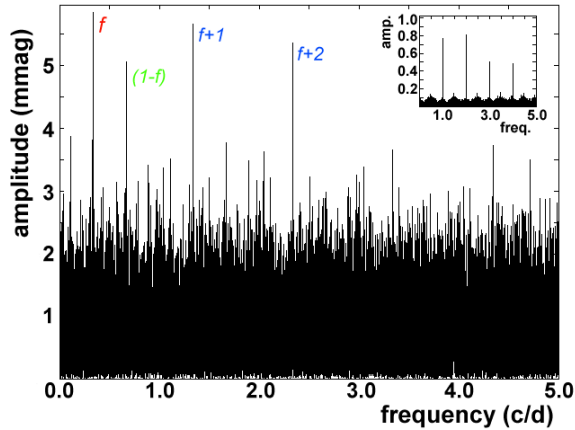}
\caption{Frequency spectrum of HD 115226, based on ASAS-3 data. The main frequency and its most prominent daily aliases are identified. The inset shows the spectral window dominated by daily aliases.}
\label{HD115226_pa}
\end{center}
\end{figure}

We have analyzed the available ASAS-3 data for HD 115226 and detect a clear signal at a frequency of $f$ = 0.33465 c/d ($P$ = 2.9882(1) d), which lies well above the noise level (Fig. \ref{HD115226_pa}). The resulting phase plot shows a double wave which is typical of ACV variables and consistent with both magnetic poles being visible over the rotation period (Fig. \ref{indobj}). Furthermore, the derived period is in accordance with the above mentioned assumptions of \citet{2008AA...479L..29K}.
	
We have investigated Hipparcos data and confirm marginal variability with $P$ = 3.61 d, as proposed by the aformentioned authors. However, this signal is not present at all in the ASAS-3 data, which boast a much longer time base ($\sim $3000 days) and a greater number of observations (639 measurements) than Hipparcos data ($\sim 1250$ days; 116 measurements). We are therefore inclined to accept the period value derived from ASAS-3 data as real.
	
The pulsations of roAp stars are explained satisfactorily by the oblique pulsator model (\citealt{1982MNRAS.200..807K,2005MNRAS.360.1022S}). Oblique pulsation results in frequency multiplets with components that are separated by the rotation frequency of the star \citep[e.g.][]{1995PASJ...47..219T}. Thus, accurate knowledge of the rotation frequency is mandatory for the full interpretation of the frequency multiplets generated by the rotational modulation of the short period pulsations observed in roAp stars. The result of the present investigation will therefore provide a significant contribution to the deciphering of the frequency multiplet of the rapid, 10.86 min oscillations observed in HD 115226.

\subsubsection{HD 123884}  \label{HD 123884}
RM09 indicated a spectral type of B8\,He\,wk for this high-latitude early-type star, remarking, however, that the given spectral type is only an approximation and that the object apparently shows no other than hydrogen lines. Very disparate entries are found in the Catalogue of Stellar Spectral Classifications \citep{2014yCat....1.2023S} which range from B4s to A0Ib?. \citet{1988PASP..100.1084B} called attention to HD 123884 and remarked that it might not be a classical CP star but rather a post-asymptotic branch star of moderate luminosity. Querying the SIMBAD and VizieR databases, no variability studies of the object were found.

ASAS-3 data indicate a period of $P$ = 1.02101(1) d for HD 123884. While the amplitude of variability is relatively high (0.013 mag, as derived with \textsc{Period04}), the period value is close to one day; thus independent confirmation of our results are needed. However, the resulting phase plot looks convincing (Figure \ref{indobj}) and shows indications of a double-humped structure, which is typical for ACV variables. Further studies to confirm the proposed variability and unravel the underlying mechanism are encouraged.

\subsubsection{HD 149334}  \label{HD 149334}
\citet{1973AJ.....78..687B} give a spectral type of ApSr for this star, \citet{1982mcts.book.....H} –- using exactly the same objective prism plates \citep{1997A&A...327..636M} –- classified it as A9IV. \citet{1997A&A...327..636M} found a $\Delta a$ value of +0.011 mag for HD 149334, which is below their peculiarity threshold of $\geq$ 0.014 mag.

HD 149334 was identified as an eclipsing binary system of Algol-type (GCVS-type EA) by \citet{2014AcA....64..115S}, who derived a period of $P$ = 3.5444 d from ASAS $I$ band data. An analysis of ASAS-3 $V$ data confirms the results of the aforementioned investigators; because of the longer time base, the period could be refined to $P$ = 3.54420(6) d (Fig. \ref{indobj}).

As the incidence of CP2 stars in eclipsing binaries is very low (cf. section \ref{classi} and section \ref{CPD-20 1640}), HD 149334 is a potentially interesting object. However, in the light of the conflicting results mentioned above, spectroscopic confirmation of the presence of a CP2 star in the system is needed.

\subsubsection{HD 161733A = IC 4665 82}  \label{HD 161733A}
This star, which is likely a member of the open cluster IC 4665, is listed with a spectral type of B7\,He\,wk\,C in RM09. In a detailed study of the object, \citet{1977PASP...89...84L} confirm the peculiar nature of HD 161733A and conclude that the star is a B-type peculiar object whose main peculiarities are being He weak; showing an enhancement of C and, perhaps, Fe and Ti; and the presence of Mn, P, Hg, and, possibly, Sr. They also find some evidence of variation in the C \textsc{II} $\lambda$4267 and Si \textsc{II} $\lambda\lambda$4128--30 lines. We have not found a reference to a photometric variability study in the literature.

From an analysis of ASAS-3 data, we derive a photometric period of $P$ = 0.97235(1) d (Fig. \ref{indobj}). While the period value is close to one day, the significance of the detection and the amplitude of variability (0.02 mag, as derived with \textsc{Period04}) are high. HD 161733 is a visual double star, the B component being of 10th magnitude and separated from the A component by 27\arcsec. Furthermore, the RM09 catalogue indicates that HD 161733A is likely a spectroscopic binary characterized by a variable radial velocity and a supposed period of $\sim$1.8 d, which is about twice our period value. Thus, further studies are required to decide whether the derived photometric period is caused by rotational or orbital modulation.

\subsubsection{HD 186205}  \label{HD 186205}
HD 186205 was identified as being a pronounced member of the class of He-rich stars by \citet{1975PASP...87..613W}. \citet{1977A&A....60..259L} confirmed Walborn's results and derived $T\_\mathrm{eff}$ = 23,500$^{\circ}$\,K, $M/M_{\odot}$ = 12.3, $R/R_{\odot}$ = 6.0 and log($L/L_{\odot}$) = 4.0 -- typical values for a He-rich star. On the basis of their data, they did not reach conclusive results about the presence of spectral variability in this star, which is listed with a spectral type of B3\,He in the RM09 catalogue. To the best of our knowledge, HD 186205 has never been confirmed as a photometrically variable star. 

An analysis of ASAS-3 data indicates light variability with a period of $P$ = 37.28(2) d. The resulting phase plot looks convincing (Fig. \ref{indobj}), and the amplitude of variability is relatively high ($\sim$0.01 mag, as derived with \textsc{Period04}).

The time-scale of the observed variability places the star outside the period domain of typical short-period, early-type pulsators like e.g. the $\beta$ Cephei variables. On the other hand, the observed variability is reminiscent of the PV Telescopii stars (GCVS-type PVTEL), subclass 'PVTELI'. However, this class is reserved for hydrogen-deficient A or late-B supergiants, whereas HD 186205 is a confirmed dwarf star and does not show a pronounced hydrogen deficiency (\citealt{1961AnTou..28...33B,1975PASP...87..613W}).

Rotationally induced variability due to surface inhomogeneities might thus be the most promising explanation of the observed light changes, although orbital variability due to an as yet undetected companion star cannot be ruled out. Long-term spectroscopic monitoring of HD 186205 is needed to reach a final conclusion concerning the nature of the observed variability.

\section{Conclusion} \label{conclu}
We have compiled an extensive target list of magnetic chemically peculiar (CP2/4) stars from the General Catalogue of Ap, HgMn, and Am stars (RM09). In addition to that, a systematic investigation of early-type (spectral types B/A) variable stars of undetermined type in the VSX yielded additional ACV candidates, which were included in our sample.

We investigated our sample stars using publicly available observations from the ASAS-3 archive. Employing a refined methodological approach, 360 stars exhibiting photometric variability in the accuracy limit of the ASAS-3 data were found. We thereby expand on a previous sample of 323 variable stars (Paper 1) and conclude our search for new photometrically variable magnetic chemically peculiar stars in the ASAS-3 archive.

From an analysis of all available data, we conclude that our final sample is composed of 334 bona-fide ACV variables, 23 ACV candidates and three eclipsing binary systems. We present summary data, folded light curves and, if available, information from the literature for all our sample stars and discuss interesting and unusual objects in detail. In particular, we call attention to HD 66051 (V414 Pup), which was identified as an eclipsing binary system showing obvious rotational modulation of the light curve due to the presence of an ACV variable in the system.

No further statistical analyses are presented in the present paper but will be given in a future publication that will include results from the literature as well as our own investigations (Paper 1, \citealt{2015AN....336..981B}, this paper).

\acknowledgments
This project is financed by the SoMoPro II programme (3SGA5916). The research leading to these results received a financial grant from the People Programme (Marie Curie action) of the Seventh Framework Programme of the EU according to REA Grant Agreement No. 291782. The research is further co-financed by the South-Moravian Region. It was also supported by grant LG15010 (INGO II). This work reflects only the author's views so the European Union is not liable for any use that may be made of the information contained therein.


%% for the paper.  Each keyword is check against the master list during
%% copy editing.  Individual instruments can be provided in parentheses,
%% after the keyword, but they are not verified.

\vspace{5mm}
\facilities{ASAS, AAVSO}

\software{Period04}



\appendix

\setcounter{table}{1}
\begin{table*}[ht!]
\caption{Essential data for the stars identified as photometrically variable chemically peculiar stars or star candidates in the present paper.}
\label{table_master2}
\begin{center}
\begin{adjustbox}{max width=\textwidth}
\begin{tabular}{llcclcccccccccc}
\hline 
Star & ID (RM09) & $\alpha$ (J2000) & $\delta$ (J2000) & Type & Range ($V$) & Period & Epoch (HJD) & $A_{\mathrm 1}$ & $A_{\mathrm 2}$ & $\phi_{\mathrm 1}$ & $\phi_{\mathrm 2}$ & Spectral type & $(B-V)$ & $(J-K_{\mathrm s})$ \\
     &           &                  &                  &      & [mag]       & [d]    & [d]  & [mag]           & [mag]           & [rad]             & [rad]               &              & [mag]   & [mag] \\
\hline
HD 2957	          & 	760	&	00 32 44.08	&	-13 29 13.5	&	ACV	&	8.48-8.51 	&	4.6327(3)       &	4627.91(9)   &	0.011	&	0.002	&	0.846	&	0.750	&	B9 Cr Eu	&	+0.052	&	-0.017 \\
HD 3885	          &	1050	&	00 41 18.31	&	-19 51 45.9	&	ACV	&	9.79-9.81	&	1.81508(4) 	&	2910.72(4)   &	0.007	&	0.009	&	0.348	&	0.449	&	B9 Si	&	-0.090	&	-0.079 \\
HD 5823	          &	1540	&	00 59 39.29	&	-11 56 01.2	&	ACV	&	9.95-9.97	&	1.24520(2) 	&	3765.53(2)   &	0.007	&	0.004	&	0.723	&	0.646	&	F2 Sr Eu Cr	&	+0.304	&	+0.176 \\
HD 8783	          &	2110	&	01 24 00.43	&	-72 19 27.9	&	ACV	&	7.80-7.82	&	19.396(5)   	&	3794.5(4)    &	0.009	&	0.001	&	0.102	&	0.715	&	A2 Sr Eu Cr	&	+0.150	&	+0.039 \\
HD 12559	  &		&	02 03 29.36     &	+18 19 39.4	&	ACV:    &	8.41-8.44  	&	4.0358(3)       &	3338.62(8)   &	0.010	&	0.012	&	0.479	&	0.293	&	A2	&	+0.092	&	-0.034 \\
HD 16145	  &	4060	&	02 35 04.20	&	-17 17 22.3	&	ACV	&	7.64-7.67  	&	2.23766(7)      &	4525.52(4)   &	0.011	&	0.002	&	0.342	&	0.873	&	A0 Cr Sr Eu	&	+0.117	&	-0.023 \\
\end{tabular}                                                                                                                                                                   
\end{adjustbox}
\end{center}                                                                                                                                             
\end{table*}  
\setcounter{table}{1}  
\begin{table*}[ht!]
\caption{continued.}
\label{table_master2}
\begin{center}
\begin{adjustbox}{max width=\textwidth}
\begin{tabular}{llcclcccccccccc}
\hline 
Star & ID (RM09) & $\alpha$ (J2000) & $\delta$ (J2000) & Type & Range ($V$) & Period & Epoch (HJD) & $A_{\mathrm 1}$ & $A_{\mathrm 2}$ & $\phi_{\mathrm 1}$ & $\phi_{\mathrm 2}$ & Spectral type & $(B-V)$ & $(J-K_{\mathrm s})$ \\
     &           &                  &                  &      & [mag]       & [d]    & [d]  & [mag]           & [mag]           & [rad]             & [rad]               &              & [mag]   & [mag] \\
\hline
HD 62821A	&	17300	&	07 44 54.96	&	-25 00 26.2	&	ACV	&	9.56-9.62	&	3.8347(2)  	&	4509.77(8) 	&	0.029	&	0.003	&	0.697	&	0.906	&	B9 Si	&	-0.081	&	-0.083	 \\
HD 62905	&	17310	&	07 45 39.83	&	-16 58 36.6	&	ACV	&	9.66-9.68	&	16.836(4)  	&	4508.1(3)  	&	0.009	&	0.003	&	0.952	&	0.292	&	A2 Eu Cr	&	+0.057	&	-0.028	 \\
HD 64698	&	17730	&	07 53 25.00	&	-40 15 59.6	&	ACV	&	9.53-9.54	&	3.0226(1)  	&	3090.66(6) 	&	0.006	&	0.001	&	0.244	&	0.735	&	A0 Cr	&	+0.092	&	-0.004	 \\
GSC 08911-01572 &	17995	&	07 56 32.84	&	-60 53 08.9	&	ACV:	&	11.43-11.51	&	5.5805(4)  	&	4825.7(1)  	&	0.037	&	0.003	&	0.041	&	0.122	&		&		&	+0.099	 \\
HD 65208	&	17870	&	07 56 37.53	&	-28 12 28.4	&	ACV	&	9.21-9.23	&	1.14192(2) 	&	3502.55(2) 	&	0.009	&	0.000	&	0.497	&	--	&	B8 Si	&	-0.097	&	-0.107	 \\
HD 66051	&	18190	&	08 01 24.64	&	-12 47 35.8	&	EA+ACV	&	8.80-9.12	&	4.74922(1) 	&	2167.867(5)	&	0.015	&	0.004	&	0.516 &	0.695 &	A0 Si	&	-0.047	&	-0.061	 \\
\end{tabular}                                                                                                                                                                   
\end{adjustbox}
\end{center}                                                                                                                                             
\end{table*}  
\setcounter{table}{1}  
\begin{table*}[ht!]
\caption{continued.}
\label{table_master2}
\begin{center}
\begin{adjustbox}{max width=\textwidth}
\begin{tabular}{llcclcccccccccc}
\hline 
Star & ID (RM09) & $\alpha$ (J2000) & $\delta$ (J2000) & Type & Range ($V$) & Period & Epoch (HJD) & $A_{\mathrm 1}$ & $A_{\mathrm 2}$ & $\phi_{\mathrm 1}$ & $\phi_{\mathrm 2}$ & Spectral type & $(B-V)$ & $(J-K_{\mathrm s})$ \\
     &           &                  &                  &      & [mag]       & [d]    & [d]  & [mag]           & [mag]           & [rad]             & [rad]               &              & [mag]   & [mag] \\
\hline
HD 92384	&	26618	&	10 38 37.14	&	-59 07 31.1	&	ACV:	&	8.60-8.61	&	2.29382(7) 	&	4656.55(5)	&	0.007	&	0.003	&	0.885	&	0.674	&	B9	&	-0.031	&	-0.012	 \\
HD 303209	&	26727	&	10 40 12.16	&	-59 24 41.9	&	ACV:	&	10.37-10.40	&	1.60624(3) 	&	4391.90(3)	&	0.011	&	0.005	&	0.362	&	0.292	&	A0	&	+0.061	&	+0.013	 \\
HD 93500A	&	27010	&	10 46 24.96	&	-59 26 42.8	&	ACV	&	8.81-8.82	&	2.36348(7) 	&	2660.82(5)	&	0.005	&	0.000	&	0.952	&	--	&	B9 Cr Eu Sr	&	+0.068	&	-0.056	 \\
HD 94052	&	27190	&	10 50 19.82	&	-55 12 30.0	&	ACV	&	9.70-9.71	&	7.6468(7)  	&	5024.3(2) 	&	0.005	&	0.003	&	0.726	&	0.690	&	B9 Cr Eu Sr	&	+0.147	&	+0.047	 \\
HD 94681	&	27320	&	10 54 37.17	&	-58 54 25.7	&	ACV	&	9.46-9.48	&	3.5444(2)  	&	2790.52(7)	&	0.006	&	0.002	&	0.512	&	0.112	&	B9 Si	&	-0.008	&	-0.033	 \\
HD 95491	&	27510	&	11 00 35.51	&	-51 38 38.0	&	ACV	&	9.22-9.23	&	4.8538(3)  	&	2990.83(9)	&	0.004	&	0.001	&	0.634	&	0.204	&	A5 Sr Cr Eu	&	+0.257	&	+0.136	 \\
\end{tabular}                                                                                                                                                                   
\end{adjustbox}
\end{center}                                                                                                                                             
\end{table*}  
\setcounter{table}{1}  
\begin{table*}[ht!]
\caption{continued.}
\label{table_master2}
\begin{center}
\begin{adjustbox}{max width=\textwidth}
\begin{tabular}{llcclcccccccccc}
\hline 
Star & ID (RM09) & $\alpha$ (J2000) & $\delta$ (J2000) & Type & Range ($V$) & Period & Epoch (HJD) & $A_{\mathrm 1}$ & $A_{\mathrm 2}$ & $\phi_{\mathrm 1}$ & $\phi_{\mathrm 2}$ & Spectral type & $(B-V)$ & $(J-K_{\mathrm s})$ \\
     &           &                  &                  &      & [mag]       & [d]    & [d]  & [mag]           & [mag]           & [rad]             & [rad]               &              & [mag]   & [mag] \\
\hline
HD 126876	&	36120	&	14 30 03.93	&	-56 55 02.4	&	ACV	&	8.36-8.39	&	11.426(1)  	&	3804.8(2) 	&	0.013	&	0.003	&	0.842	&	0.523	&	B8 Si	&	+0.175	&	+0.049	 \\
HD 127021	&	36150	&	14 30 55.83	&	-54 32 24.4	&	ACV	&	9.80-9.81	&	2.01366(5) 	&	4514.87(4)	&	0.007	&	0.004	&	0.698	&	0.492	&	B9 Sr Cr	&	+0.274	&	+0.138	 \\
HD 127224	&	36200	&	14 32 56.61	&	-65 08 18.2	&	ACV	&	9.83-9.86	&	6.5031(5)  	&	4669.5(1) 	&	0.012	&	0.003	&	0.693	&	0.733	&	B8 Si	&	+0.234	&	+0.055	 \\
HD 127924	&	36350	&	14 36 01.53	&	-56 27 36.5	&	ACV:	&	9.22-9.24	&	3.5487(2)  	&	2868.59(7)	&	0.008	&	0.001	&	0.460	&	0.831	&	B8	&	+0.099	&	+0.090	 \\
HD 127957	&	36360	&	14 36 33.23	&	-59 17 28.4	&	ACV	&	9.00-9.01	&	3.2801(1)  	&	3798.82(7)	&	0.008	&	0.001	&	0.579	&	0.852	&	B9 Si	&	-0.002	&	+0.014	 \\
HD 128226	&	36440	&	14 38 05.24	&	-59 09 47.3	&	ACV	&	10.02-10.04	&	2.48567(8) 	&	4715.58(5)	&	0.010	&	0.004	&	0.806	&	0.418	&	B9 Si	&	+0.063	&	+0.023	 \\
\end{tabular}                                                                                                                                                                   
\end{adjustbox}
\end{center}                                                                                                                                             
\end{table*}  
\setcounter{table}{1}  
\begin{table*}[ht!]
\caption{continued.}
\label{table_master2}
\begin{center}
\begin{adjustbox}{max width=\textwidth}
\begin{tabular}{llcclcccccccccc}
\hline 
Star & ID (RM09) & $\alpha$ (J2000) & $\delta$ (J2000) & Type & Range ($V$) & Period & Epoch (HJD) & $A_{\mathrm 1}$ & $A_{\mathrm 2}$ & $\phi_{\mathrm 1}$ & $\phi_{\mathrm 2}$ & Spectral type & $(B-V)$ & $(J-K_{\mathrm s})$ \\
     &           &                  &                  &      & [mag]       & [d]    & [d]  & [mag]           & [mag]           & [rad]             & [rad]               &              & [mag]   & [mag] \\
\hline
HD 150137	&	42460	&	16 41 54.33	&	-54 07 36.8	&	ACV	&	8.82-8.86	&	5.1512(4)  	&	4181.9(1) 	&	0.016	&	0.001	&	0.899	&	0.200	&	B9 Si	&	+0.110	&	+0.016	 \\
CD-59 6262	&	42470	&	16 43 16.14	&	-59 59 50.2	&	ACV	&	9.88-9.91	&	0.844887(9)	&	2503.58(2)	&	0.015	&	0.003	&	0.632	&	0.193	&	Si	&	-0.019	&	-0.022	 \\
CD-60 6444	&	42720	&	16 49 00.54	&	-60 28 00.8	&	ACV	&	9.60-9.62	&	4.4673(3)  	&	3583.73(9)	&	0.010	&	0.007	&	0.381	&	0.375	&	B9 Si	&	+0.011	&	-0.059	 \\
HD 152099	&	43040	&	16 53 43.44	&	-47 24 48.4	&	ACV	&	10.18-10.20	&	3.2074(1)  	&	4276.45(6)	&	0.011	&	0.002	&	0.379	&	0.332	&	A0 Si	&	+0.230	&	+0.067	 \\
HD 152137	&	43070	&	16 54 49.91	&	-57 25 04.1	&	ACV	&	9.66-9.68	&	1.33044(2) 	&	4756.51(3)	&	0.006	&	0.002	&	0.561	&	0.543	&	A2 Sr Cr Eu	&	+0.161	&	+0.108	 \\
HD 152834	&	43260	&	16 56 58.02	&	-26 57 05.6	&	ACV	&	8.84-8.87	&	4.2928(3)  	&	3602.32(9)	&	0.013	&	0.000	&	0.606	&	--	&	A0 Si	&	+0.017	&	-0.006	 \\
\end{tabular}                                                                                                                                                                   
\end{adjustbox}
\end{center}                                                                                                                                             
\end{table*}  
\setcounter{table}{1}  
\begin{table*}[ht!]
\caption{continued.}
\label{table_master2}
\begin{center}
\begin{adjustbox}{max width=\textwidth}
\begin{tabular}{llcclcccccccccc}
\hline 
Star & ID (RM09) & $\alpha$ (J2000) & $\delta$ (J2000) & Type & Range ($V$) & Period & Epoch (HJD) & $A_{\mathrm 1}$ & $A_{\mathrm 2}$ & $\phi_{\mathrm 1}$ & $\phi_{\mathrm 2}$ & Spectral type & $(B-V)$ & $(J-K_{\mathrm s})$ \\
     &           &                  &                  &      & [mag]       & [d]    & [d]  & [mag]           & [mag]           & [rad]             & [rad]               &              & [mag]   & [mag] \\
\hline
HD 181549	&	50280	&	19 22 46.22	&	-41 18 29.8	&	ACV:	&	10.30-10.32	&	22.871(6)  	&	3851.5(5) 	&	0.006	&	0.003	&	0.330	&	0.974	&	A0	&	+0.066	&	+0.000	 \\
HD 231382	&	50540	&	19 25 43.69	&	+16 14 21.4	&	ACV	&	9.30-9.34	&	4.4773(3)  	&	4684.69(9)	&	0.014	&	0.003	&	0.351	&	0.343	&	B9 Si	&	+0.291	&	+0.109	 \\
HD 183735	&	50740	&	19 30 50.66	&	+01 02 42.2	&	ACV	&	9.84-9.86	&	2.23029(6) 	&	3164.72(4)	&	0.010	&	0.002	&	0.781	&	0.822	&	A2 Sr Cr Eu	&	+0.323	&	+0.071	 \\
HD 184539	&	50910	&	19 34 04.53	&	+24 35 16.4	&	ACV	&	8.75-8.78	&	2.57198(9) 	&	4607.77(5)	&	0.013	&	0.003	&	0.249	&	0.669	&	A1 Si Sr	&	+0.061	&	-0.005	 \\
HD 186205	&	51340	&	19 42 37.85	&	+09 13 39.0	&	ACV:	&	8.57-8.58	&	37.28(2)   	&	3527.7(7) 	&	0.006	&	0.001	&	0.086	&	0.589	&	B3 He	&	+0.057	&	-0.072	 \\
HD 353530	&	51450	&	19 45 06.18	&	+14 08 09.5	&	ACV	&	9.98-10.00	&	2.08664(6) 	&	2737.91(4)	&	0.009	&	0.002	&	0.648	&	0.544	&	A0 Si	&	+0.014	&	-0.054	 \\
\end{tabular}                                                                                                                                                                   
\end{adjustbox}
\end{center}                                                                                                                                             
\end{table*}

\setcounter{table}{2}
\begin{sidewaystable*}
\begin{center}
\caption{Relevant information on single objects from the literature and miscellaneous remarks. An asterisk in column 7 ('Remarks/comments') denotes stars whose status as chemically peculiar objects is doubtful according to RM09. The following abbreviations are employed in column 7: R12 = \citet{2012MNRAS.427.2917R}; W12 = \citet{2012MNRAS.420..757W}.}
\label{table_remarks2}
\begin{adjustbox}{max width=\textwidth}
\scriptsize{
\begin{tabular}{llccccl}
\hline
\hline
Star &  Var. desig. & Var. type &       Period (d)  &   Period (d)      &       Reference       &       Remarks/comments        \\
     &  Literature       & Literature     & Literature  & This work   &           & Literature \\      
\hline
HD 2957		&				&		&						&	4.6327(3)  	&		&	Null result for roAp pulsations \citep{1994MNRAS.271..129M}. \\
HD 3885		&	NSV 15149		&	VAR	&						&	1.81508(4) 	&	\citet{1982AAS...49..427W}	&	\\
HD 5823		&				&		&						&	1.24520(2) 	&		&	Null result for roAp pulsations \citep{2013MNRAS.431.2808K}. \\
HD 8783		&	SMC V2339		&	VAR:	&	21(or 1.05)? (GCVS)			&	19.396(5)  	&	GCVS	&	Non-member of the SMC according to the GCVS. The given period \\
		& 				& 		& 						& 	           	& 		& values are derived from variations of the peculiarity index $\Delta$a. \\
HD 12559	&	HIP 9602		&	VAR	&	2.01833 (VSX);  2.01837 (R12)		&	4.0358(3)  	&	\citet{2002MNRAS.331...45K}	&	R12: SPB/ACV (prob: 0.24/0.51) \\
\hline
\end{tabular}
}
\end{adjustbox}
\end{center}  
\end{sidewaystable*}

\setcounter{table}{2}

\begin{sidewaystable*}
\centering
\caption{continued.}
\label{table_remarks2}
\scriptsize{
\begin{adjustbox}{max width=\textwidth}
\begin{tabular}{llccccl}
\hline
\hline
Star &  Var. desig. & Var. type &       Period (d)  &   Period (d)      &       Reference       &       Remarks/comments        \\
     &  Literature       & Literature     & Literature  & This work   &           & Literature \\      
\hline
HD 106204	&	NSV 19303	&	VAR	&						&	4.3572(3)  	&	\citet{1982AAS...48..503R}	&		 \\
BD+01 2668	&			&		&						&	0.98410(1) 	&		&	*	 \\
HD 109300	&			&		&						&	2.63189(9) 	&		&	*	 \\
HD 110072	&			&		&						&	22.225(6)  	&		&	Null result for roAp pulsations \citep[e.g.][]{1994MNRAS.271..129M}. \\
		& 			& 		& 						& 	           	& & $v\,sin\,i$ = 3.3 $\pm$ 0.5; $T_\mathrm{eff}$ = 7300 K; double-lined binary with \\
		& 			& 		& 						& 	           	& & peculiar lines of e.g. Nd, Pr and Eu \citep{2008MNRAS.385.1402F}. \\
\hline
\end{tabular}
\end{adjustbox}
}
\end{sidewaystable*}
\setcounter{table}{2}
\begin{sidewaystable*}[ht!]
\centering
\caption{continued.}
\label{table_remarks2}
\scriptsize{
\begin{adjustbox}{max width=\textwidth}
\begin{tabular}{llccccl}
\hline
\hline
Star &  Var. desig. & Var. type &       Period (d)  &   Period (d)      &       Reference       &       Remarks/comments        \\
     &  Literature       & Literature     & Literature  & This work   &           & Literature \\      
\hline
HD 173657	&	HIP 92215	&	VAR	&	1.93720 (VSX)						&	1.93789(5) 	&	\citet{2002MNRAS.331...45K}	&		 \\
HD 173361	&			&		&								&	0.831981(9)	&		&	*	 \\
HD 181549	&			&		&								&	22.871(6)  	&		&	*	 \\
HD 193325	&			&		&	3.153d? (RM09)						&	3.1529(1)  	&		&		 \\
HD 198918	&	NP Del		&	ELL:	&	0.777813 (VSX); 0.388866 \citep{2011MNRAS.414.2602D}	&	0.777809(7)	&	\citet{1997AA...323L..61V}	&		 \\
HD 200199	&			&		&								&	1.03512(1) 	&		&	*	 \\
HD 205013	&	NSV 25655	&	VAR	&	1.57594 (VSX)						&	1.57624(3) 	&	\citet{1991TrSht..63....1K}	&		 \\
HD 209605A	&			&		&								&	7.8132(7)  	&		&	Null result for roAp pulsations \citep{1994MNRAS.271..129M}.	 \\
\hline
\end{tabular}
\end{adjustbox}
}
\end{sidewaystable*}

\setcounter{figure}{7}
\begin{figure*}
\begin{center}
\includegraphics[width=1.0\textwidth,natwidth=1446,natheight=1710]{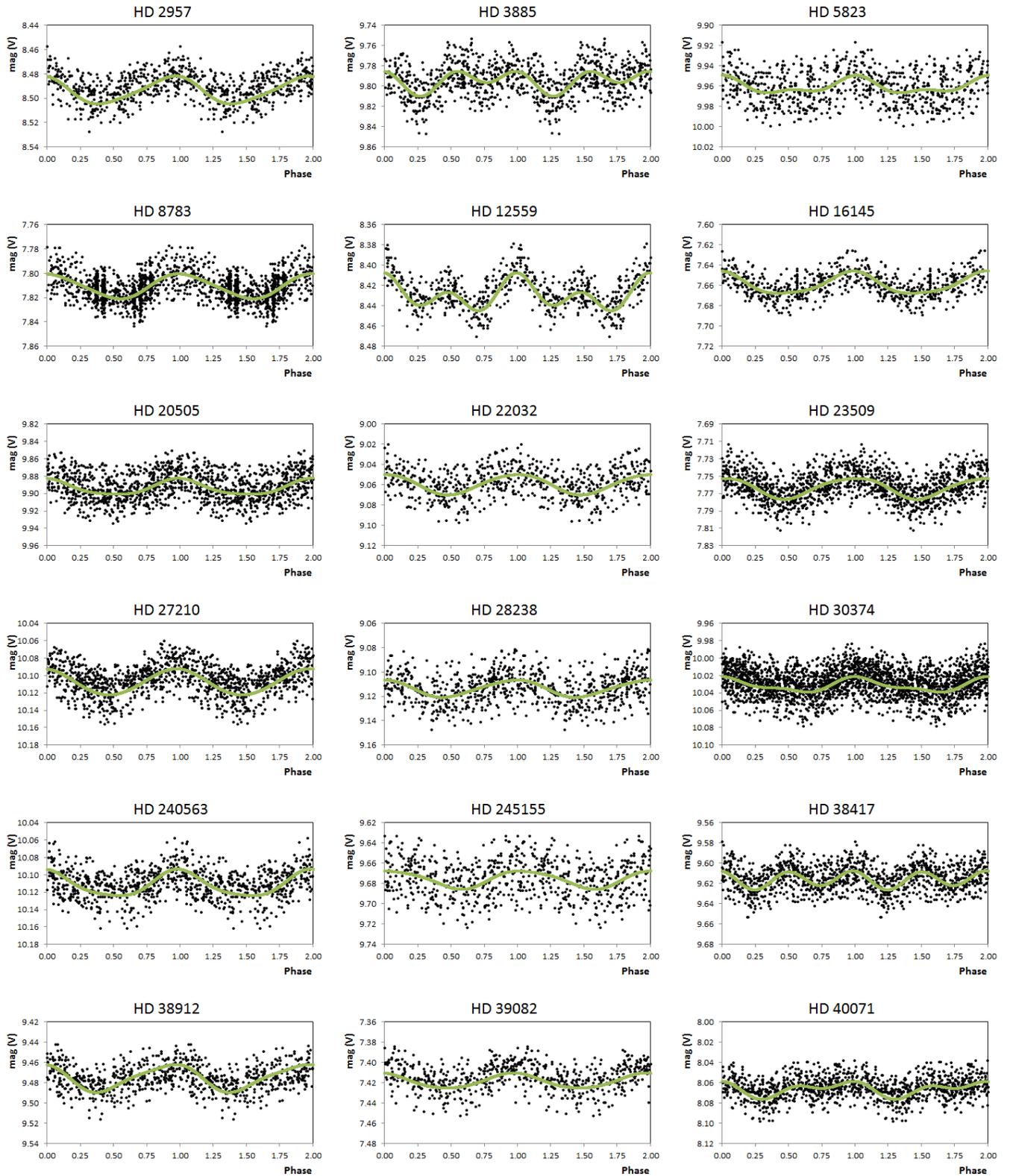}
\caption{The light curves of all objects, folded with the periods listed in Table \ref{table_master}. The fit curves corresponding to the light curve parameters
given in Table \ref{table_master} are indicated by the solid lines. The complete figure is available from the authors.}
\end{center}
\end{figure*}




\end{document}